\documentclass[osajnl,showpacs,preprint]{revtex4}
\usepackage{amsmath}
\usepackage{graphicx} 
\usepackage{dcolumn}  
\usepackage{bm}       

\begin{document}

\author{Albrecht Haase}
\email{haase@physi.uni-heidelberg.de}
\affiliation{%
Physikalisches Institut, Universit\"at Heidelberg, 69120 Heidelberg,
Germany}
\author{Bj\"orn Hessmo}
\email{hessmo@physi.uni-heidelberg.de}
\affiliation{%
Physikalisches Institut, Universit\"at Heidelberg, 69120 Heidelberg,
Germany}
\author{J\"org Schmiedmayer}

\email{schmiedmayer@physi.uni-heidelberg.de}
\affiliation{%
Physikalisches Institut, Universit\"at Heidelberg, 69120 Heidelberg,
Germany}
\title{Detecting magnetically guided atoms with an optical cavity}
\begin{abstract}
We show that a low finesse cavity can be efficient for detecting
neutral atoms. The low finesse can be compensated for by decreasing
the mode waist of the cavity. We have used a near concentric resonator
with a beam waist of 12$\mu$m and a finesse of only 1100 to detect
magnetically guided Rb atoms with a detection sensitivity of 0.1
atom in the mode volume. For future experiments on single atom detection and cavity
QED applications, it should be very beneficial to use
miniaturized optical resonator integrated on atom chips.
\end{abstract}

\maketitle

It is highly desirable to detect atoms with high efficiency and good
spatial resolution both for fundamental physical experiments\cite{Ber94} and 
for applications in quantum information processing\cite{DiV00}.
This task is usually accomplished by using high finesse cavities,
where the photons interact strongly with the atoms. In such
experiments it has been possible to monitor the motion of single
atoms inside a high finesse ($\mathcal{F}>2\times 10^5$) cavity\cite{Mab99}. 
By performing feedback to the optical beam
passing through the resonator it has been possible to control the
atomic motion inside the cavity to form bound atom-photon states.
This also requires a high finesse ($\mathcal{F}>4\times
10^5$) resonator\cite{Pin00}.

In this article we wish to explore the possibilities to 
perform atom detection using optical cavities with moderate 
finesse\cite{Hor03}. 
The main result is that the finesse $\mathcal{F}$ is not the most important
aspect of cavity assisted detection schemes, equally important is
the ratio between the atomic absorption cross section
$\sigma_{\mathrm{atom}}=3\lambda^2/2\pi$ and the beam cross section
inside the cavity $A=\frac{\pi}{4}w_0^2$.

For each round trip the photon is absorbed with a probability
$\sigma_{\mathrm{atom}}/A$. A figure of merit for absorption
inside the cavity is therefore
\begin{equation}
C_1=\frac{\mathcal{F}}{2\pi}\frac{\sigma_{\mathrm{atom}}}{A}.
\label{eq:Purcell}
\end{equation}
where $\frac{\mathcal{F}}{2\pi}$ is the number of round trips for a
photon. This quantity is identical to the cooperativity
parameter $C_1=\frac{g_0^2}{2\kappa\gamma}$, which
relates the time scales of the coherent dynamics of the coupled
system $g_0^{-1}$ to the time scales of incoherent decays of cavity
field $\kappa^{-1}$ and atomic excitation $\gamma^{-1}$. This is
also related to the Purcell factor $\eta=2C_1$  that
determines the enhancement of the spontaneous emission rate into the
cavity mode over the free space value\cite{Ber94,Pur46}. 
Looking at Eqn.~(\ref{eq:Purcell}) one clearly sees that a reduced cavity mode waist
can compensate for a small cavity finesse.

In this spirit it has been proposed by Horak \emph{et al.} that a single
atom detection in low finesse can be achieved by strongly focussing
the cavity mode\cite{Hor03}. When the cooperativity parameter is smaller than
one and the atomic saturation is low the signal-to-noise ratio for a
single atom detection becomes
$$
S=\sqrt{j_\mathrm{in}\tau}\frac{\kappa_T}{\kappa}C_1,
$$
where $j_\mathrm{in}$ is the incident photon flux, $\tau$ the
measurement interval, $\kappa_T$ the mirror transmission rate, and
$\kappa$ the overall cavity decay rate\cite{Hor03}. For a fixed
measurement time an increased signal-to-noise ratio can be obtained
by increasing the cooperativity parameter. This can be done by
increasing the cavity finesse, or by decreasing the beam waist. 
Here we explore the latter case, when the beam cross section is
reduced.

To achieve this we use a nearly concentric cavity geometry. Our
cavity was formed by two identical mirrors with radius of curvature
$R$ separated by a distance $L$. The beam waist $w_0$ in the
cavity center is given by
$w_{0}^2=\frac{\lambda}{2\pi}\sqrt{L(2R-L)}$. The concentric
geometry occurs when the mirror separation $L$ approaches the value
$2R$. The waist size $w_0$ becomes small but the beam size on the
cavity mirrors $w^2=\frac{R\lambda}{\pi}\sqrt{\frac{L}{2R-L}}$
becomes large as one approaches the concentric limit.

A large mirror spot size requires very uniform mirrors as deviations
from a spherical mirror shape will lower the optical finesse
drastically. Furthermore, as the concentric point is approached, the
cavity also becomes extremely sensitive to misalignments and
vibrations.  For more details on this cavity we refer to 
Siegman\cite{Sie86}.


Table~\ref{tab:finesses} summarizes parameters for nearly concentric
cavities and shows that it is not stringently necessary to have
$C_1>1$  to detect the presence of a single atom within the cavity
mode.

These considerations all concern the coupling of a single atom to a
cavity. This can be generalized to the many atom case by introducing
a many-atom cooperativity parameter $C=N_\mathrm{eff}C_1$, where
$N_\mathrm{eff}$ is an effective number of atoms in the cavity mode\cite{Haa05}, 
which takes into account the spatial dependency of the
coupling constant $g(\vec{r})=g_0 \psi(\vec{r})$, given by the
cavity mode function $\psi(\vec{r})$, and the atomic density
distribution $\rho(\vec{r})$. The fraction of the total atom number
$N$ which is maximally coupled to the cavity mode is given by the
overlap integral of both functions
\begin{equation}
N_\mathrm{eff}=N \int d^3r\,\rho(\vec{r})|\psi(\vec{r})|^2.
\label{eq:Neff}
\end{equation}
The absorptive and dispersive effect of the atoms on the cavity
amplitude\cite{Hor03} scale linearly with this effective atom
number as long as the atomic saturation is low.

\begin{figure}[h]
    \includegraphics[width=0.7\textwidth]{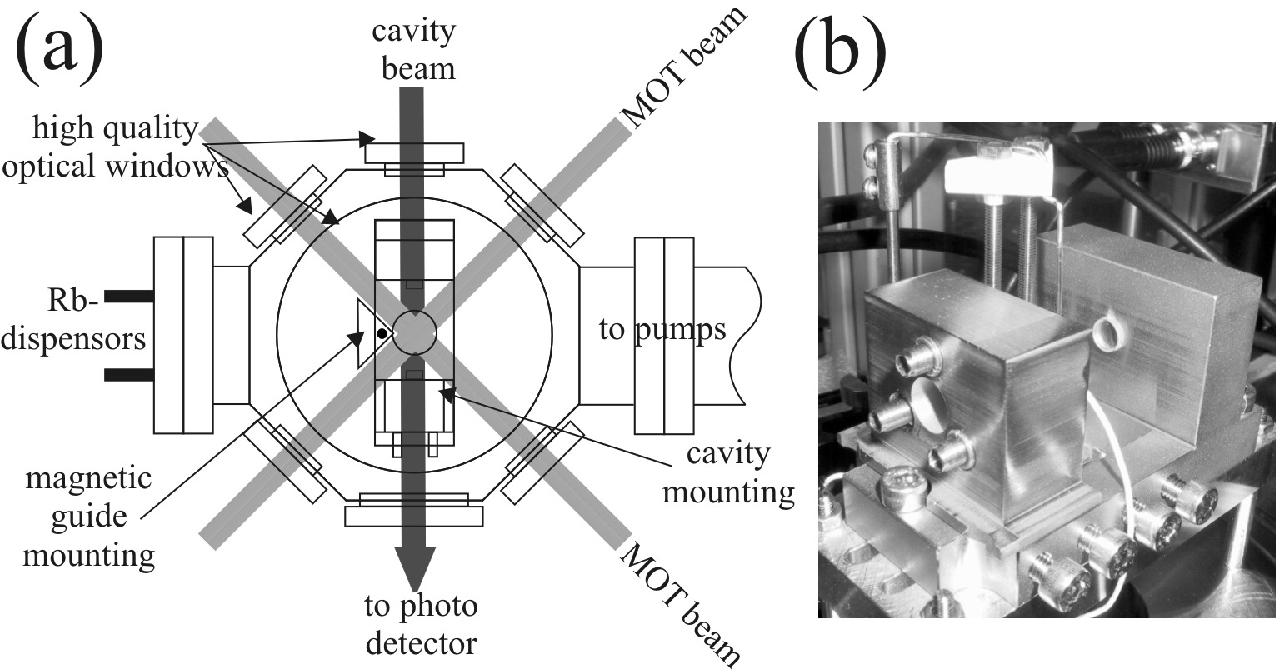}
    \centering
    \caption{}\label{fig:setup}
\end{figure}

To explore atom detection with low finesse cavities experimentally,
we built a magneto-optical trap (MOT) for ${}^{85}$Rb atoms
approximately 20mm above the cavity center (see Fig.
\ref{fig:setup}). It contained $\sim 10^{7}$ atoms at a temperature of 35$\mu$K. From the
MOT we proceeded in two different ways. Either we switched off
the trap completely and monitored the atomic cloud as it fell freely
through the cavity, or we transfered the atoms to a magnetic wire guide
that channeled the atoms through the cavity\cite{Den99}. The magnetic guide was
formed by a current-carrying wire, attached to the cavity mounting
in vertical direction (see Fig. \ref{fig:setup}b) and a homogeneous
magnetic bias field in the direction of the optical axis of the cavity.
In this configuration a two dimensional quadrupole guide was formed.
The depth, confinement, and position was controlled by
varying the wire current and the magnetic bias field\cite{Den99,Haa01}.

\begin{figure}[h]
    \includegraphics[width=0.7\textwidth]{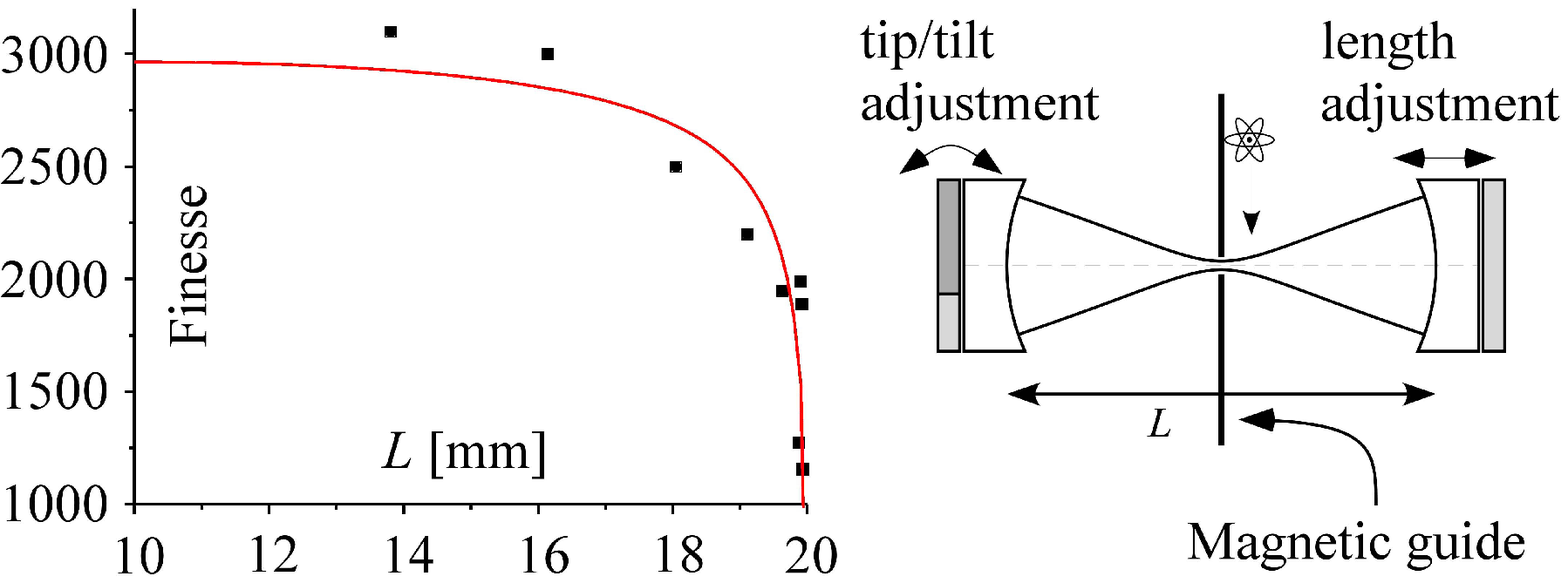}
    \centering
    \caption{}
    \label{fig:cavity}
\end{figure}

To keep the cavity aligned, we mounted one of the mirrors on a
piezoelectric tripod that allowed us to adjust the optical axis of
the cavity. This mirror was aligned to keep the TEM${}_{00}$ mode
centered on the cavity axis. The second mirror was mounted on a
translating piezoelectric stage for wavelength tuning. Feedback to
this piezo actuator was generated using the Pound-Drever-Hall
technique\cite{Dre83} to lock the cavity on the laser beam passing
through the cavity.
%
%
Figure \ref{fig:cavity} illustrates how the cavity finesse was
reduced as the concentric point was approached for our cavity. We used
two mirrors with $R=10$mm and transmission $T=10^{-3}$. For a mirror
separation far less than the concentric limit these mirrors yielded a
finesse of 3000. This finesse dropped to 1100 when the separation was 
70$\mu$m from the concentric point as discussed in the introduction. 
The cavity mode waist was 12.1$\mu$m for this separation.

We monitored the light intensity transmitted through with an amplified
photodiode for high light intensities or with a photomultiplier tube
(PMT) for low light intensities. The PMT provided a near shot-noise
limited detection. The low-noise electronic amplification limited 
the detection bandwidth to 20kHz. The main source of technical noise 
in our setup was due to mechanical vibrations of the vacuum chamber 
that held the cavity.
%

\begin{figure}[h]
    \includegraphics[width=0.7\textwidth]{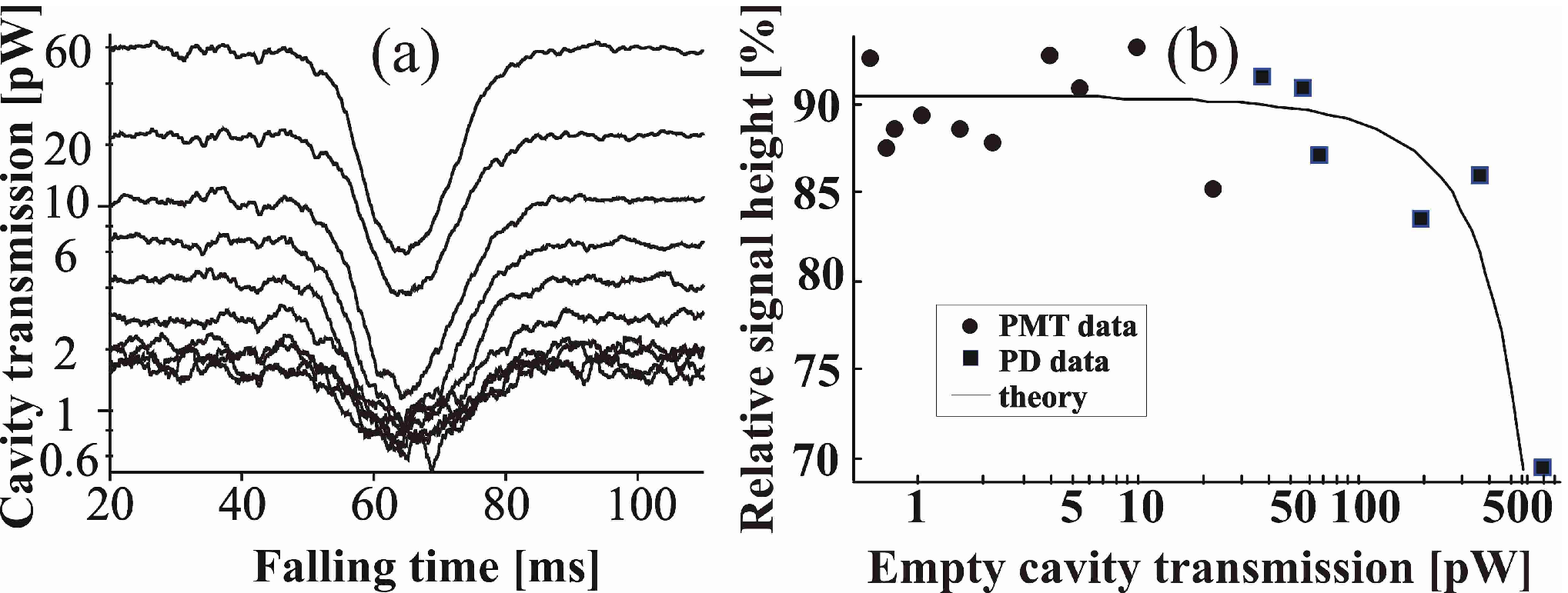}
    \centering
    \caption{}
    \label{fig:freefall}
\end{figure}

The drop in the cavity transmission signal from freely falling atoms
is plotted in Fig. \ref{fig:freefall}a). Different curves manifest
different pump powers corresponding to empty cavity transmissions
between 1pW and 60pW. The atom number
in the MOT is $1.5\times10^7$, the signal drops by 90$\%$ as long as
the atomic transition is not saturated (Fig. \ref{fig:freefall}b).
Fitting this data with the theoretical model\cite{Haa05}, one
obtains an effective atom number $N_\mathrm{eff}=2.5\pm0.5$.
This was consistent with an independent atom number measurement 
based on florescence imaging. To explore the sensitivity limit of 
the cavity detector, the atom number in the MOT was successively 
reduced until the signal drop due to the atoms was overshadowed by 
the noise. 

When the MOT contained $3.5\times10^5$ atoms this produced a signal
drop of approximately $10\%$. We consider this to be the resolution
limit. A theoretical fit results in an effective atom number of
$N_\mathrm{eff}=0.1\pm0.05$. 
%
%

\begin{figure}[h]
    \includegraphics[width=0.7\textwidth]{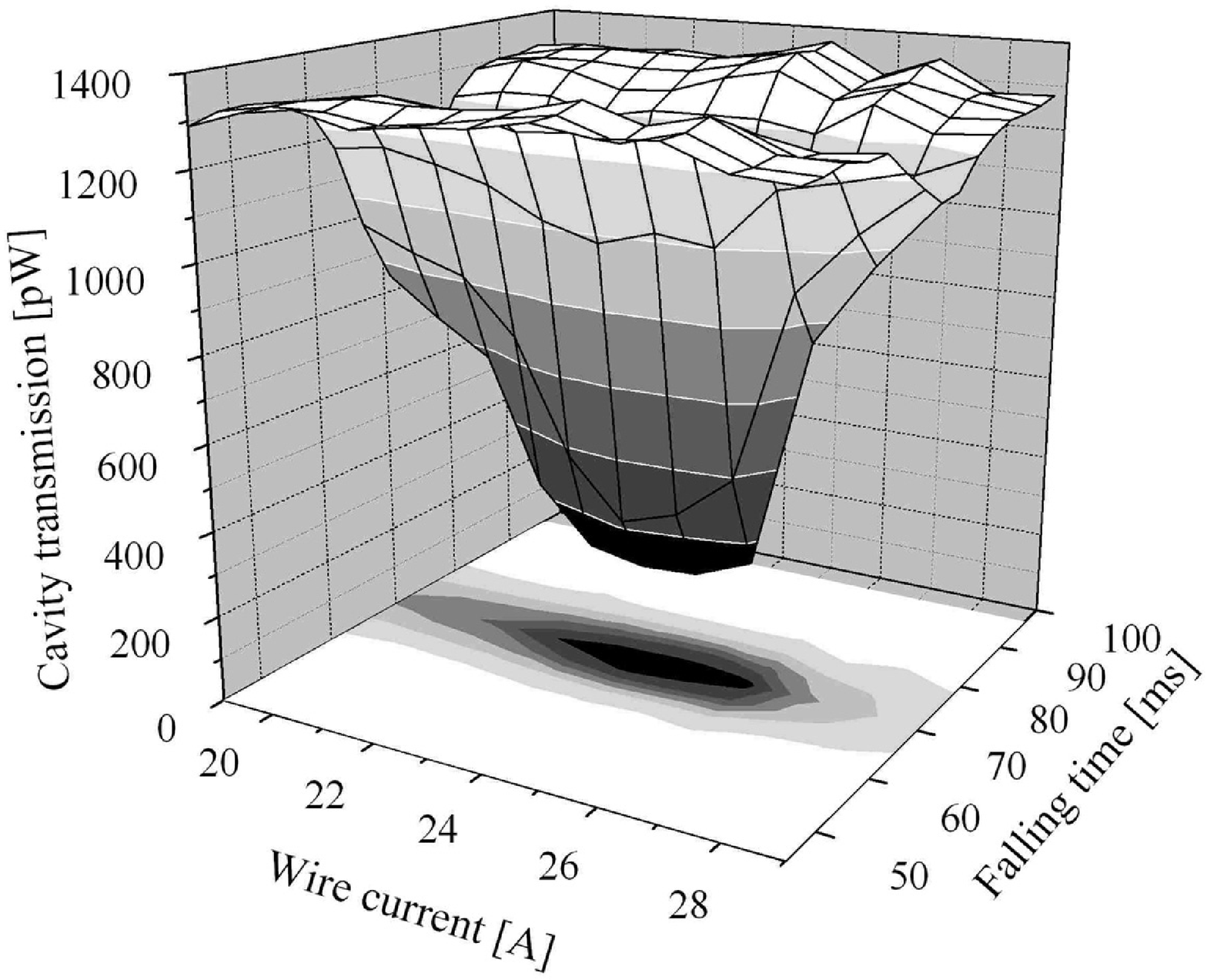}
    \centering
    \caption{}
\label{fig:guide}
\end{figure}

As a next step, atoms were magnetically guided to the cavity center 
using the wire guide (see Fig.~\ref{fig:setup}). By changing the
current in the guiding wire the overlap between the atoms and the
cavity mode could be adjusted. In Fig.~\ref{fig:guide} we plot the
cavity transmission as the position of the magnetic guide is varied
across the cavity mode. As the atomic overlap with the cavity mode
was increased, we observed a increased drop in cavity transmission.
From the duration of the transmission drop the temperature of the
guided atoms could be determined to be 25$\mu$K.

The density distribution for the atoms was much larger than the 
Rayleigh volume of the cavity, consequently it was not possible to 
distinguish individual atoms in the guide using our low finesse cavity. 
This cavity would however show a detectable change in the transmission 
signal if a single atom would cross the region of maximum 
coupling as $N_\mathrm{eff}$ can be as small as $0.1$. The precision 
in the positioning can be improved using magnetic microtraps, e.g. 
produced by atom chip surface traps\cite{Fol02}. On the atom chip 
one can also build small integrated cavities\cite{Liu05} with mode waists as small 
as 2$\mu$m. This relaxes the requirements on the finesse 
even further\cite{Hor03}. To achieve the same detection sensitivity 
with a beam waist of 2$\mu$m a finesse of 40 is enough.

To conclude, we have illustrated that it is possible to detect 
magnetically guided atoms using a low finesse cavity with small 
mode waist. The small waist allowed us to detect atoms with high 
sensitivity, as illustrated in Fig.~\ref{fig:freefall}. 
We also show that high spatial resolution can be achieved. 
We demonstrate this by detecting magnetically guided atoms, as 
illustrated in Fig.~\ref{fig:guide}. A natural development would 
be to miniaturize the cavity even further and integrate 
it on an atom chip\cite{Hor03,Arm03,Lev04,Mok04,Liu05}.

We gratefully acknowledge valuable discussions with P. Horak, T.
Fernholz, and M. Wilzbach. Funding was provided by Landesstiftung
Baden-W\"urttemberg, Forschungsprogramm
Quanteninformationsverarbeitung and the EU-program MRTN-CT-2003-50532.


\pagebreak
\begin{table}
\begin{tabular}{|c|c|c|c|c|c|c|}\hline
$L$ & $w_0$ & $\mathcal{F}$ & $g_0$ & $\kappa$ & $C_1$ & $S_\mathrm{max}$\\\hline
[mm] & $[\mu$m]&  &  $2\pi\times$[MHz] & $2\pi\times$[MHz] & & \\\hline
19.0 & 23.3 & 1000 & 1.6 & 3.9 & 0.1 & 2.8\\
19.93 & 12.1 & 1000 & 3.0 & 3.8 & 0.4 & 6.1\\
19.99 & 7.5 & 1000 & 4.9 & 3.8 & 1.1 & 13.2\\
19.99 & 7.5 & 300 & 4.9 & 12.5 & 0.3 & 5.3\\\hline
\end{tabular}
  \centering
  \caption{}\label{tab:finesses}
\end{table}

\pagebreak

\pagebreak
\begin{itemize}
    \item\textbf{Caption Tab. I:}
    In the table the cooperativity parameters and maximum
    signal-to-noise ratios for various cavity geometries close to the
    concentric cavity limit at $L=2R=20$mm are listed. The atomic decay
    rate is that of the rubidium D2 transition and $\tau=10\mu s$.
    
    \item\textbf{Caption Fig. 1:} (a) Schematic drawing of the experimental chamber which
    contains a magneto-optical trap, the quasi-concentric cavity and
    a magnetic wire guide. (b) Picture of the cavity mounting including
    guiding wire.
    
    \item\textbf{Caption Fig. 2:} a) The finesse of the cavity decreases as the concentric
    point is approached. The curve is calculated from the cavity geometry 
    and mirror specifications. b) Schematic description of the cavity.
    The two mirrors with 10mm radius of curvature are mounted on
    piezo electric actuators for alignment. One of the mirrors is can
    be tilted to keep the optical axis of the cavity fixed. The other
    mirror can be translated for frequency tuning. Atoms can
    be magnetically guided through the cavity.
    
    \item\textbf{Caption Fig. 3:} (a) Cavity transmission signal for atoms
    dropped from a MOT. Different curves correspond to cavity pump powers from 1-60pW.
    The signal has been averaged over 2.5ms for better visualization.
    (b) Relative drop of the signal due to the atoms in (a). The circles (squares) come from measurements with a PMT (photodiode) for
    different light intensities. 
    The black line is calculated numerically.
    
    \item\textbf{Caption Fig. 4:} Cavity transmission signal from atoms being magnetically
    guided through the cavity mode. The position of the potential minimum is
    linearly dependent on the wire current.
\end{itemize}

\end{document}